\documentclass[twocolumn,aps,pra,showpacs]{revtex4}

\begin{document}
\title{Rotationally invariant proof of Bell's theorem without inequalities}
\author{Ad\'{a}n Cabello}
\email{adan@us.es}
\affiliation{Departamento de F\'{\i}sica
Aplicada II, Universidad de Sevilla, 41012 Sevilla, Spain}
\date{\today}


\begin{abstract}
The singlet state of two spin-${3 \over 2}$ particles allows a
proof of Bell's theorem without inequalities with two
distinguishing features: any local observable can be regarded as
an Einstein-Podolsky-Rosen element of reality, and the
contradiction with local realism occurs not only for some specific
local observables but for any rotation whereof.
\end{abstract}


\maketitle


\section{Introduction}


Mermin's version \cite{Mermin90a,Mermin90b,Mermin90c} of the
Greenberger-Horne-Zeilinger (GHZ) proof
\cite{GHZ89,GHSZ90,Greenberger02} has been considered ``the most
simple, surprising, and convincing'' \cite{Vaidman01} proof of
Bell's discovery \cite{Bell64} of the fact that
Einstein-Podolsky-Rosen (EPR) ``elements of reality'' \cite{EPR35}
are incompatible with quantum mechanics (QM). On the other hand,
Hardy's argument of ``nonlocality without inequalities''
\cite{Hardy93} has been considered ``the best version of Bell's
theorem'' \cite{Mermin95}. Besides their beauty and simplicity,
however, both proofs lack of one of the distinguishing features of
the original proof by Bell: rotational invariance. Bell's proof is
based on Bohm's version \cite{Bohm51} of the EPR experiment using
the singlet state of two spin-${1 \over 2}$ particles. According
to EPR, only {\em ``if, without in any way disturbing a system, we
can predict with certainty (i.e., with probability equal to unity)
the value of a physical quantity, then there exists an element of
physical reality corresponding to this physical quantity''}
\cite{EPR35}. Therefore, for the singlet state, and for any
maximally entangled state of two spin-$s$ particles, {\em any}
spin observable on each particle can be regarded as an element of
reality. In contrast, for any GHZ state of three or more spin-$s$
particles \cite{Cabello01} or for any Hardy state (i.e., an
entangled but not maximally entangled pure state) of two spin-$s$
particles \cite{CN92}, only the results of some local observables
can be predicted with certainty from spacelike separated
measurements. Therefore, not all local observables can be regarded
as elements of reality. Moreover, while in Bell's proof the
disagreement between elements of reality and QM occurs for a
continuous range of local observables, both in the GHZ and Hardy's
proofs the algebraic contradictions between EPR elements of
reality and QM appear only for a specific set of local
observables, but vanishes for any other choice of observables.

A natural question is then, would it be possible to prove Bell's
theorem, without using inequalities, on a physical system in which
(a) any local observable satisfies EPR's criterion for elements of
reality, and (b) the contradiction between QM and elements of
reality appears not for some specific local observables but for a
continuous range of them?

It can be proved that there does not exist a rotationally
invariant GHZ state of three or more particles, and it is easy to
see that Hardy states are not rotationally invariant. However, a
proof of Bell's theorem without inequalities which fulfills the
above requirements is described in the following section.


\section{Proof without inequalities}


Let us consider two observers, Alice and Bob, in two distant
regions. Each of them receives a spin-${3 \over 2}$ particle
belonging to a pair initially prepared in the singlet state which,
using the standard choice for the matrices $S_x$ (symmetric and
real) and $S_y$ (antisymmetric and pure imaginary) \cite{Peres95},
representing the spin along the $x$ and $y$ directions, can be
expressed as
\begin{eqnarray}
\left|{\psi } \right\rangle & = & {1 \over 2} ( \left|3/2, -3/2
\right\rangle - \left|1/2,-1/2 \right\rangle \nonumber \\ & &
+ \left| -1/2,1/2 \right\rangle - \left| -3/2,3/2 \right\rangle ).
\label{singlet32}
\end{eqnarray}
The notation is the following: $|3/2,-3/2 \rangle =
|3/2\rangle_{\rm A} \otimes |-3/2 \rangle_{\rm B}$, where
$|3/2\rangle_{\rm A}$ is the eigenstate with eigenvalue $3/2$
($\hbar =1$) of the spin along the $z$ direction of Alice's
particle. We shall choose $\langle 3/2|=(1,0,0,0)$, $\langle
1/2|=(0,1,0,0)$, $\langle -1/2|=(0,0,1,0)$, and $\langle
-3/2|=(0,0,0,1)$. The singlet state~(\ref{singlet32}) is
rotationally invariant, which means that, if we act on both
particles with the tensor product of two equal rotation operators,
the result will be to reproduce the same state (within a possible
phase factor). A method for preparing optical analogs of the
singlet state of two $n$-dimensional systems for an arbitrary high
$n$ has been recently described and has been experimentally
implemented for low $n$ \cite{LHB01,HLB02}.

It can be easily seen that, in the state~(\ref{singlet32}), {\em
every} local spin observable satisfies EPR's criterion for
elements of reality: its value can be predicted with certainty by
a spacelike separated measurement on the other particle.
Specifically, let us consider the local observables represented by
the operators
\begin{eqnarray}
D & = & \left( {\matrix{ 1&{}&{}&{}\cr {}&{1}&{}&{}\cr
{}&{}&{-1}&{}\cr {}&{}&{}&{-1}\cr }} \right),
\label{A} \\
d & = & \left( {\matrix{ 1&{}&{}&{}\cr {}&{-1}&{}&{}\cr
{}&{}&{1}&{}\cr {}&{}&{}&{-1}\cr }} \right),
\label{a} \\
U & = & \left( {\matrix{ {0}&{0}&{1}&{0}\cr {0}&{0}&{0}&{1}\cr
{1}&{0}&{0}&{0}\cr {0}&{1}&{0}&{0}\cr }} \right),
\label{B} \\
u & = & \left( {\matrix{ {0}&{1}&{0}&{0}\cr {1}&{0}&{0}&{0}\cr
{0}&{0}&{0}&{1}\cr {0}&{0}&{1}&{0}\cr }} \right),
\label{b}
\end{eqnarray}
and also the observables represented by the operators $Dd$, $Du$,
$Ud$, $Uu$.

As can be easily checked, in the singlet state~(\ref{singlet32}),
the result $r_{\rm A}(D)$, either $-1$ or $1$, of Alice's
measurement of the observable $D$ on her particle and the result
$r_{\rm B}(D)$ of Bob's measurement of $D$ on his particle are
opposite. Moreover, it can be easily checked that, in the singlet
state~(\ref{singlet32}), the following correlations between
Alice's and Bob's results would occur:
\begin{eqnarray}
r_{\rm A}(D) & = & -r_{\rm B}(D),
\label{one} \\
r_{\rm A}(d) & = & -r_{\rm B}(d),
\label{two} \\
r_{\rm A}(U) & = & r_{\rm B}(U),
\label{three} \\
r_{\rm A}(u) & = & -r_{\rm B}(u),
\label{four} \\
r_{\rm A}(Dd) & = & r_{\rm B}(D) r_{\rm B}(d),
\label{five} \\
r_{\rm A}(Uu) & = & -r_{\rm B}(U) r_{\rm B}(u),
\label{six} \\
r_{\rm A}(D) r_{\rm A}(u) & = & r_{\rm B}(Du),
\label{seven} \\
r_{\rm A}(U) r_{\rm A}(d) & = & -r_{\rm B}(Ud),
\label{eight} \\
r_{\rm A}(Dd) r_{\rm A}(Uu) & = & r_{\rm B}(Du) r_{\rm B}(Ud).
\label{nine}
\end{eqnarray}
Let us show that any of the $12$ local observables (six per
particle) appearing in Eqs.~(\ref{one})--(\ref{nine}) satisfies
EPR's criterion for elements of reality and thus possesses a
preexisting result, either $-1$ or $1$, which is revealed when the
corresponding measurement is performed, and is not altered when
another compatible observable is measured. In particular, let us
examine Eqs.~(\ref{five})--(\ref{nine}), since each of them
involves two elements of reality of the same particle. Let us
take, for instance, Eq.~(\ref{five}). Following EPR and using
Eq.~(\ref{one}), a measurement of $D$ on Bob's particle would
reveal a preexisting element of reality $r_{\rm B}(D)$. Likewise,
following EPR and using Eq.~(\ref{two}), a measurement of $d$ on
Bob's particle would reveal a preexisting element of reality
$r_{\rm B}(d)$. However, it could happen that the measurement of
$D$ could alter the preexisting element of reality $r_{\rm B}(d)$.
How can we guarantee that a previous measurement of $D$ will not
affect the result of a subsequent measurement of (the compatible
observable) $d$? We can guarantee it by invoking EPR's criterion
for elements of reality: since, by means of a spacelike separated
measurement on his particle, Alice can predict with certainty
$r_{\rm B}(d)$ using Eq.~(\ref{two}), regardless of whether or not
Bob has measured $D$ before measuring $d$, then, following EPR, we
conclude that Bob's measurement of $D$ does not change $r_{\rm
B}(d)$. If $d$ was an element of reality for Bob's particle, a
measurement of $D$ on Bob's particle does not alter the
preexisting element of reality of $d$. The same reasoning applies
whenever a pair of compatible local observables is measured on the
same particle, as in Eqs.~(\ref{five})--(\ref{nine}).

The proof of Bell's theorem of incompatibility between elements of
reality and QM comes from the fact that it is impossible to assign
preexisting results, either $-1$ or $1$, to the $12$ local
observables in such a way that satisfies the predictions of QM
given by Eqs.~(\ref{one})--(\ref{nine}). This can be checked in
the following manner: if we take the product of
Eqs.~(\ref{one})--(\ref{nine}), each result (either $-1$ or $1$)
appears {\em twice} in each side, since each operator appears
twice in each side. Therefore, the product of the left-hand sides
must be $1$, while the product of the right-hand sides must be
$-1$. We, therefore, conclude that any physical theory in which
the notion of EPR elements of reality makes sense cannot reproduce
the predictions of QM for the singlet state of two spin-${3 \over
2}$ particles given by Eqs.~(\ref{one})--(\ref{nine}).


\section{How to measure the local observables}


Let us now describe how to measure the local observables involved
in the proof on a spin-${3 \over 2}$ particle. Observable $D$ can
be measured by measuring the spin along the $z$ direction using a
Stern-Gerlach device: if the result of this measurement is
$S_z=3/2$ or $S_z=1/2$, then $r(D)=1$; if the result is $S_z=-1/2$
or $S_z=-3/2$, then $r(D)=-1$. Likewise, observable $d$ can be
measured by measuring the spin along the $z$ direction; if the
result is $S_z=3/2$ or $S_z=-1/2$, then $r(d)=1$; if the result is
$S_z=1/2$ or $S_z=-3/2$, then $r(d)=-1$. Therefore, a joint
measurement of $D$ and $d$, like the one Bob needs to check
Eq.~(\ref{five}), is equivalent to a measurement of the spin along
the $z$ direction on Bob's particle; if the result is $S_z=3/2$,
then $r_{\rm B} (D)=r_{\rm B} (d)=1$; if $S_z=1/2$, then $r_{\rm
B} (D)=-r_{\rm B} (d)=1$; if $S_z=-1/2$, then $r_{\rm B}
(D)=-r_{\rm B} (d)=-1$; and if $S_z=-3/2$, then $r_{\rm B}
(D)=r_{\rm B} (d)=-1$. Any observable represented by a diagonal
operator can be measured using this method. The recipe for
measuring the other local observables involved in the proof is not
as easy. In the case of a single spin-${3 \over 2}$ particle,
these observables require generalized Stern-Gerlach devices as
described in Ref.~\cite{SW80}. Other possibility is to implement
optical analogs of these observables by using the method described
in Ref.~\cite{RZBB94}.


\section{Rotational invariance}


The most relevant feature of the proof introduced above is that it
is rotationally invariant: the contradiction between elements of
reality and QM occurs not only for the particular set of
observables $\{D_{\rm A},D_{\rm B},d_{\rm A},\ldots\}$ used above,
but for any set of observables $\{{\cal R} D_{\rm A},{\cal R}
D_{\rm B},{\cal R} d_{\rm A},\ldots\}$, where ${\cal R} D_{\rm A}$
is the physical observable obtained by applying a rotation ${\cal
R}$ to the device for measuring the observable $D_{\rm A}$. As can
be easily checked, in the singlet state~(\ref{singlet32}),
\begin{eqnarray}
r_{\rm A}({\cal R}D) & = & -r_{\rm B}({\cal R}D),
\label{oner} \\
r_{\rm A}({\cal R}d) & = & -r_{\rm B}({\cal R}d),
\label{twor} \\
r_{\rm A}({\cal R}U) & = & r_{\rm B}({\cal R}U),
\label{threer} \\
r_{\rm A}({\cal R}u) & = & -r_{\rm B}({\cal R}u),
\label{fourr} \\
r_{\rm A}({\cal R}Dd) & = & r_{\rm B}({\cal R}D) r_{\rm B}({\cal R}d),
\label{fiver} \\
r_{\rm A}({\cal R}Uu) & = & -r_{\rm B}({\cal R}U) r_{\rm B}({\cal R}u),
\label{sixr} \\
r_{\rm A}({\cal R}D) r_{\rm A}({\cal R}u) & = & r_{\rm B}({\cal R}Du),
\label{sevenr} \\
r_{\rm A}({\cal R}U) r_{\rm A}({\cal R}d) & = & -r_{\rm B}({\cal R}Ud),
\label{eightr} \\
r_{\rm A}({\cal R}Dd) r_{\rm A}({\cal R}Uu) & = & r_{\rm B}({\cal R}Du) r_{\rm B}({\cal R}Ud).
\label{niner}
\end{eqnarray}
Therefore, once Alice and Bob have found a set of observables
leading to a contradiction between elements of reality and QM,
then {\em any} common rotation of the local observables will lead
to a similar contradiction, without rotating the source of
entangled pairs.


\section{Conclusions}


To sum up, we have presented a proof of Bell's theorem for the
singlet state of two spin-${3 \over 2}$ particles which combines
the simplicity of the GHZ proof with the symmetry of the original
proof by Bell, in which any local spin observable can be regarded
as an Einstein-Podolsky-Rosen element of reality, and in which the
contradiction between elements of reality and QM occurs for a
continuous range of settings.


\begin{acknowledgments}
I thank J. L. Cereceda and C. Serra for comments, and the Spanish
Ministerio de Ciencia y Tecnolog\'{\i}a Grant No.~BFM2002-02815
and the Junta de Andaluc\'{\i}a Grant No.~FQM-239 for support.
\end{acknowledgments}


\end{document}